\documentclass[sigconf,screen]{acmart}

\usepackage{tabularx}
\usepackage{multirow}
\usepackage{booktabs}
\usepackage{array}
\usepackage{diagbox}
\usepackage{longtable}
\usepackage{makecell}
\usepackage{subcaption}
\usepackage{float}
\usepackage{placeins}
\usepackage{longtable}
\usepackage{stfloats}

\AtBeginDocument{%
  }

\copyrightyear{2025}
\acmYear{2025}

\begin{document}

\title{TA-GNN: Physics Inspired Time-Agnostic Graph Neural Network for Finger Motion Prediction}

\author{Tinghui Li}
\affiliation{%
 \institution{University of Sydney}
 \country{Australia}
 \postcode{2008}}
\email{tinghui.li@sydney.edu.au}
\orcid{0009-0009-3121-8997}

\author{Pamuditha Somarathne}
\affiliation{%
 \institution{University of Sydney}
 \country{Australia}}
\email{pamuditha.somarathne@sydney.edu.au}
\orcid{}

\author{Zhanna Sarsenbayeva}
\affiliation{%
 \institution{University of Sydney}
 \country{Australia}}
\email{zhanna.sarsenbayeva@sydney.edu.au}
\orcid{0000-0002-1247-6036}

\author{Anusha Withana}
\affiliation{%
 \institution{University of Sydney}
 \country{Australia}}
\email{anusha.withana@sydney.edu.au}
\orcid{0000-0001-6587-1278}

\renewcommand{\shortauthors}{Li et al.}

\begin{abstract}

Continuous prediction of finger joint movement using historical joint positions/rotations is vital in a multitude of applications, especially related to virtual reality, computer graphics, robotics, and rehabilitation. However, finger motions are highly articulated with multiple degrees of freedom, making them significantly harder to model and predict. To address this challenge, we propose a physics-inspired time-agnostic graph neural network (TA-GNN) to accurately predict human finger motions. The proposed encoder comprises a kinematic feature extractor to generate filtered velocity and acceleration and a physics-based encoder that follows linear kinematics. The model is designed to be prediction-time-agnostic so that it can seamlessly provide continuous predictions. The graph-based decoder for learning the topological motion between finger joints is designed to address the higher degree articulation of fingers. We show the superiority of our model performance in virtual reality context. This novel approach enhances finger tracking without additional sensors, enabling predictive interactions such as haptic re-targeting and improving predictive rendering quality.

\end{abstract}



\begin{teaserfigure}
    \centering
    \includegraphics[width=\textwidth]{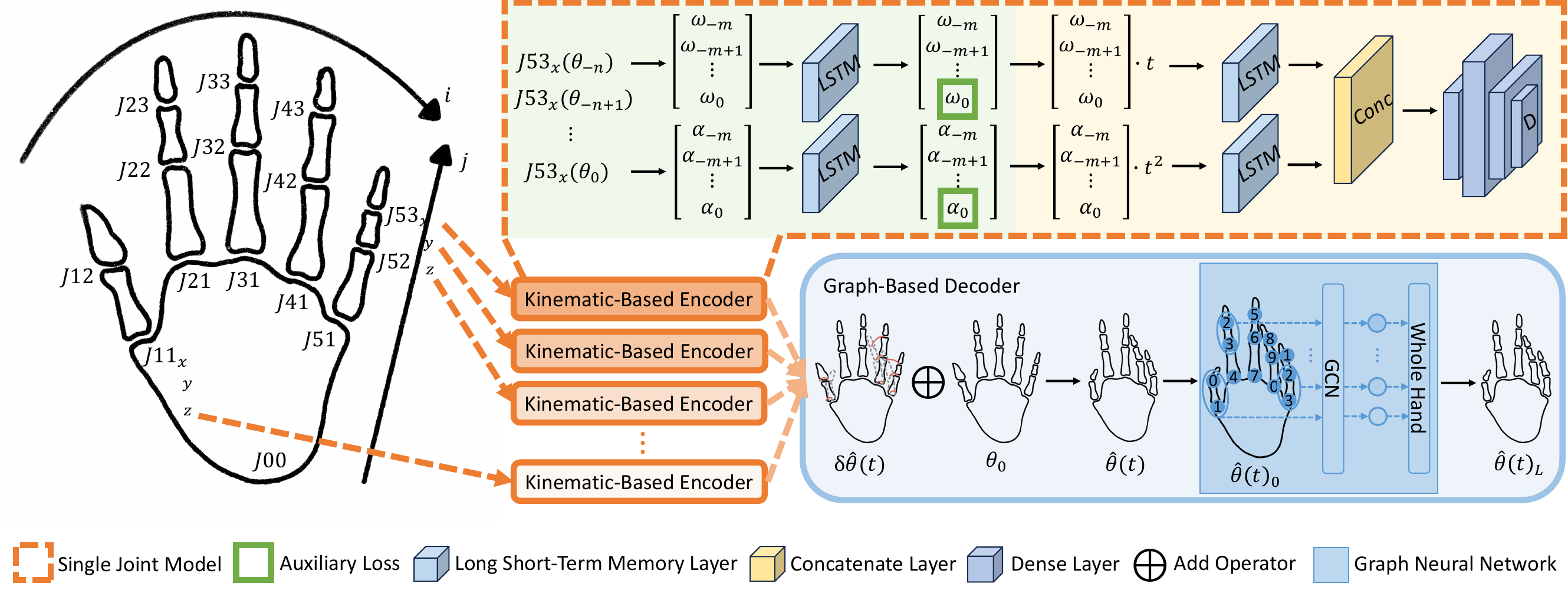}
    \caption{The architecture of TA-GNN, which consists of a kinematics-based encoder and a graph-based decoder. The kinematic-based encoder (orange dash-box) consists of the kinematic feature extractor (green background) and the physics-based encoder (yellow background), which is a joint-axis-specific model. The graph-based decoder (blue background) generates the future motion as a whole hand. The auxiliary loss calculation is applied in the green box. }
    \label{fig:model}
\end{teaserfigure}


\maketitle

\section{Introduction}
\label{sec:intro}

Human motion prediction has many applications in different fields, such as human-computer interaction \cite{koppula2015anticipating,gui2018teaching}, autonomous driving \cite{chen20203d}, human tracking \cite{alahi2016social,gupta20143d,bhattacharyya2018long}, and virtual reality (VR) \cite{gamage2021so}. This method allows systems to understand human behavior and generate realistic motion. It also gives people the opportunity to interact with the systems in a more intuitive and immersive way. Human motion prediction can encompass both full-body movements \cite{li2020dynamic} as well as specific body parts, including the upper limbs \cite{ren2019deep}, lower limbs \cite{zhu2023semg}, and fingers \cite{vangi2023enhancing,li2017preliminary}. 

Particularly in VR, hand and finger interactions have become a widely used method for interacting with virtual objects \cite{tanaka2017immersive}. Even though finger-tracking plays a vital role in this interaction method, finger-tracking technologies still experience significant errors and delays due to occlusions, lagging, incorrect graphics rendering, and discrete motion blur \cite{abdlkarim2022methodological}, thus leading to poor user experiences. In fields like limb or whole-body motion tracking, motion prediction models have been able to improve on such tracking issues \cite{gamage2021so}. However, the hands and fingers have many degrees of freedom for movement and the nature of finger movement is highly articulated, making the modelling of interactions between hands and fingers more challenging \cite{chan2007hand}. Accurate and continuous finger motion prediction models can assist in pre-rendering of graphics, anticipating user behavior, and real-time correction of errors. 

Specifically, continuous prediction is vital in the above applications as it can provide flexibility in predictions \cite{jayaraman2018time}. Unlike models that only predict motion at certain time periods, a continuous time model can predict how a finger moves continuously without additional training steps. This is crucial for smooth movement predictions, especially when a user interacts with virtual objects \cite{gamage2021so} with their fingers. Since the learned weights of such a model need not be changed with prediction time, we call them time-agnostic models. 

The set of linear kinematic equations is one of the most powerful tools in physics of free body motion, capable of accurate time-agnostic predictions. Drawing inspiration from this, we design our model architecture to capture the kinematics of finger joint movement. However, each finger joint does not follow these equations individually due to the forces exerted by muscles, changes in human intentions, and interconnections between the joints. This results in the need for non-linear prediction models. In order to extract the kinematic features of each joint (i.e. angular velocity and angular acceleration as shown in Figure~\ref{fig:model} green boxes) from historical data, we use the kinematic feature extractor module, which is shown in Figure~\ref{fig:model} green background. We use auxiliary loss functions to ensure the accurate calculation of these kinematic features. To enforce kinematic constraints and encode temporal features of each joint we follow this with the time-agnostic physics-based encoder module (Figure~\ref{fig:model} yellow background). For learning the spatial domain interconnections between joints, we apply a graph convolutional network as the decoder (Figure~\ref{fig:model} blue background) to extract the topological movement information across different finger joints. Given that each finger joint maintains an associative relationship with its parent joint, we set up the edge connections of the decoder to learn and predict how the fingers move in a realistic way. Combining these three components, we derive a time-agnostic deep-learning model called TA-GNN, which is able to learn both the temporal and topological information within and across each finger joint. The architecture of TA-GNN is illustrated in Figure~\ref{fig:model}.

The numerous large-scale datasets \cite{ionescu2013human3, punnakkal2021babel, mahmood2019amass} available for whole-body motion prediction have been one of the key contributors to the popularity and rapid advancements in the field. However, in comparison finger motion datasets are limited in number, especially for VR applications \cite{jarque2020large, roda2022studying, yang2023fully}. Some of the available datasets face constraints in accuracy and generalization as they have been collected using data gloves with strain gauges \cite{roda2022problems}. We conducted a user study to collect a novel finger motion capture dataset called VRHands dataset in VR context. The experiments conducted on the VRHands dataset as well as existing Re:InterHand \cite{moon2023reinterhand} dataset show that TA-GNN provides significant improvements compared to baseline. We also show that TA-GNN outperforms the methods introduced previously by Chen et al. \cite{chan2007hand} and Yang et al. \cite{yang2023fully}.

The main contributions of the paper can be summarized as: 

\begin{itemize}

    \item We propose TA-GNN -- a physics-inspired time-agnostic graph neural network -- based on historical kinematic features and topological movement information. TA-GNN consists of a kinematic feature extractor, a physics-based encoder, and a graph-based decoder;
    
    \item The experiment results show the proposed TA-GNN significantly outperform baseline models and capable of accurately predicting future movement for time-agnostic instances;

    \item We propose a new direction for improving finger tracking without additional sensors in virtual reality, enabling predictive interactions such as haptic re-targeting and improving predictive rendering quality.
    

\end{itemize}


\section{Related work}\label{sec:relatedwork}

\subsection{Human motion prediction}
One of the primary issues with the current hand-tracking technology is the delay in recognizing hand movements, which can cause users to feel disconnected from the virtual environment \cite{pyasik2020visual} and reduce user performance \cite{li2025encumbrance}. Human motion prediction has been used to improve lagging issues in different applications such as human tracking \cite{bhattacharyya2018long} and VR \cite{gamage2021so,yang2023fully}. Researchers have been working on developing computer vision-based movement recognition systems that use inexpensive technologies. In recent years, many deep learning-based approaches such as graph neural networks, have been proposed to achieve whole-body skeleton-based motion prediction \cite{guo2023back}. Furthermore, trajectory predictions have been proposed to predict the movement of the upper limbs \cite{gamage2021so}, where the upper body is divided into joints and bones for prediction. Gamage et al. proposed a model based on VR movements, where prediction can be user- and activity-independent \cite{gamage2021so}. However, as fingers move fast and in a smaller volume than comparatively larger body parts like limbs, finger motion needs finer predictions than the whole body. Therefore, we aim to develop a model that is specifically designed around finger motion. Additionally, there are methods that use surface electromyography (sEMG) signals \cite{vangi2023enhancing}, ultra-sound signals \cite{zadok2023towards}, and stereo-electroencephalographic (sEEG) signals \cite{li2017preliminary} to predict hand motion. However, these sensors are not commonly available and, therefore, unsuitable for widespread VR applications.

\subsection{Graph based methods for motion prediction}
Using graph-based learning to capture spatial relations between the body joints has been shown to be effective in predicting motions \cite{li2021multiscale,li2020dynamic,li2021symbiotic}. Li et al. proposed a dynamic multiscale graph neural network that uses the relations between different body components \cite{li2021multiscale}. Similarly, Dang et al. proposed a novel multi-scale residual graph convolution network for end-to-end human pose prediction tasks \cite{dang2021msr}. Bae et al. discovered unknown low-dimensional skeletal relationships that can effectively represent motions \cite{bae2022neural}. These approaches show that graph-based motion prediction could achieve high precision, leveraging the inherent connectivity and relationships among the finger joints. Besides, other areas including computer vision could also achieve high prediction results~\cite{sun2024awf,hu2025improvingcognitivediagnosticspathology}. 

\subsection{Finger motion capture datasets}
The existing research mostly focuses on whole-body movement from motion capture (MoCap) systems. However, large-scale motion capture datasets such as Human3.6M \cite{ionescu2013human3} do not contain finger movement data, particularly collected from VR scenarios. Additionally, some existing datasets on finger motion capture, have been collected through sensor gloves such as 22-sensor CyberGlove data glove \cite{jarque2020large,roda2022studying} and flexible intelligent glove \cite{yang2023fully}. While data gloves with strain gauges, like the CyberGlove, are commonly used for recording hand kinematics, they often face issues with fitting different individuals and accurately recording movements \cite{roda2022problems}. Hand motion datasets such as InterHand2.6M \cite{moon2020interhand}, Re:InterHand \cite{moon2023reinterhand} contain hand videos as well as corresponding joint annotations. However, they have been annotated semi-automatically with help from human annotators. While these datasets are useful for developing and evaluating prediction models, the data annotation method used is not feasible in large-scale applications such as everyday VR applications. Hence, we collected a novel finger rotation dataset to supplement the already available finger motion datasets by conducting a study that focused especially in VR context. This dataset will be made publically available to facilitate further research.


\section{Methodology}\label{sec:method}

In this section, we introduce the physics-inspired kinematic equation that is the basis of our architecture. Then we explain the reasoning and details behind the three components of the model: the kinematic feature extractor, the physics-based encoder, and the graph-based decoder. Finally, we present the loss function that we used to optimize the model during training.

\subsection{Physics-inspired motion capture}

\begin{figure}[!ht]
    \centering
    \includegraphics[scale=0.22]{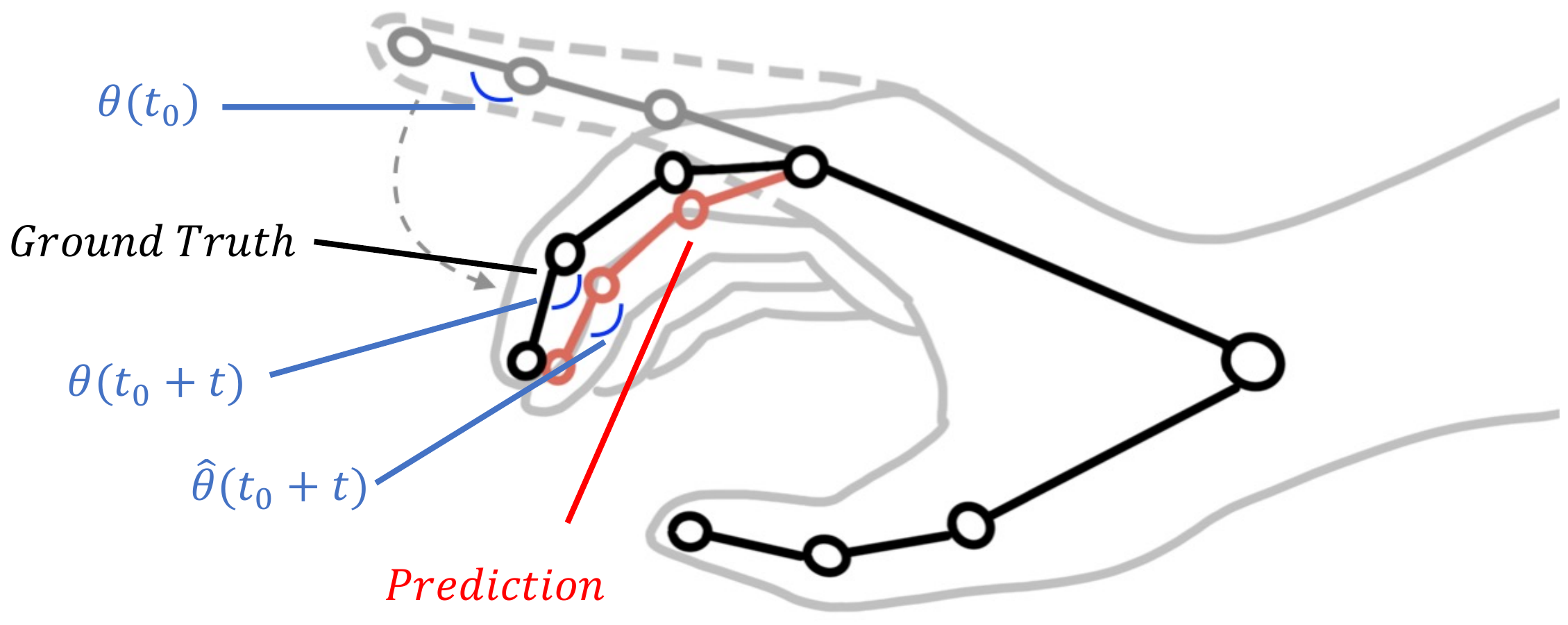}
    \caption{The joint rotation prediction process. The gray joints show the previous actual position of the finger. The red joints show the prediction ($\hat{\theta}(t_0 + t)$) for $t$ milliseconds into the future from a given time $t_0$, while the black joints show the actual position of the finger ($\theta(t_0 + t)$) in the future. }
    \label{fig:angle}
\end{figure}

Classical kinematics is a foundational branch of mechanics dedicated to the study of the motion of objects. It offers an intricate framework that permits the analysis of the behavior of free-moving entities relative to a specified frame of reference. Building on top of that, Flash and Hogan specify that human arm movements follow a minimum jerk trajectory where jerk refers to the third-order derivative of the motion \cite{flash1985mjt}. This implies that for human arm motion, the derivatives above some order $k$ should be zero. In our work, we employ the classical kinematic equation, as denoted by Equation~\ref{eq:kinematic_general} with up to $k^{th}$ order derivatives. In our experiments for finger motion, we notice $k=2$ as a suitable trade-off between containing important motion information and avoiding high-frequency noise distortions. We denote these two derivatives as angular velocity ($\omega$) and angular acceleration ($\alpha$) respectively. This is also sufficient for capturing vital motion dynamics. Here, $\theta$ represents the joint rotation and $t$ stands for prediction time, which is also shown in Figure~\ref{fig:angle}. 

\begin{equation}
    \theta (t_0 + t) = \sum_{i=0}^{k} \frac{1}{i!} \frac{d^i \theta(t_0)}{dt^i} t^i
    \label{eq:kinematic_general}
\end{equation}

As the prediction interval grows larger, we encounter a problem when using classical kinematics equations for human motion. The constant weights do not allow the flexibility in prediction to cater to changes in motion due to joint interactions, human intentions, and muscle adjustments for feedback from the brain. To address this limitation, we integrate neural networks to model learnable non-linear functions directly into classical kinematics. This integration of modern computing techniques with classical methods represents a significant advancement, helping us overcome the limitations by adding a layer of adaptability and precision to our predictions. 

\subsection{Kinematic feature extractor}\label{ssec:kfe}

\begin{figure}[!ht]
    \centerline{\includegraphics[scale = 1]{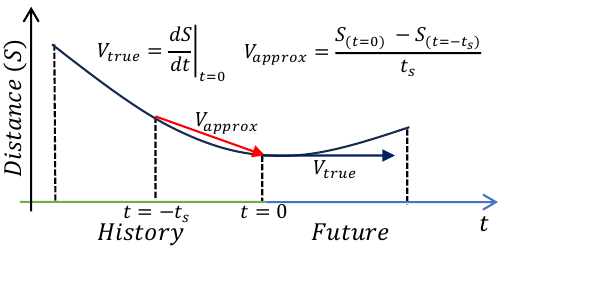}}
    \caption{The difference that LSTM layers used to correct derivatives in the kinematic feature extractor. $V_{approx}$ (red arrow) represents the approximate value and $V_{true}$ (blue arrow) represents the actual value. $t_s$ is the sampling interval. }
    \label{fig:derivative}
\end{figure}

The functionality of the kinematic feature extractor is to calculate the angular velocity and angular acceleration to be used in the kinematic equation. However, due to the instability when recognizing hands in virtual environment, historical angle data contain noise. To address this issue, we use a moving-window average filter to filter the historical data. This guarantees the exclusion of future data from input while reducing the impact of anomalous data. Approximations for angular velocity and angular acceleration are calculated from these filtered angles. As shown in Figure~\ref{fig:derivative}, it is impossible to calculate accurate kinematic features from historical data alone. Therefore, the extractor consists of LSTM layers that minimize the deviation of these derivatives as shown in Figure~\ref{fig:model}. 

We use auxiliary loss functions as expressed in Equation~\ref{eq:velocity_loss} and Equation~\ref{eq:acceleration_loss} to enforce correct angular velocity and angular acceleration calculation. These compare the predicted angular velocity ($\hat{\omega}_0$) and angular acceleration ($\hat{\alpha}_0$) (highlighted with green boxes in Figure~\ref{fig:model}) with angular velocity ($\omega_{g0} = G_v(\theta_0)$) and angular acceleration ($\alpha_{g0} = G_a(\theta_0)$) calculated using a non-causal Gaussian filter. Each joint ($j \in J$) has its own kinematic feature extractor, as the scale of movement varies between joints. 

\begin{equation}
    L_\omega = \frac{1}{m} \sum_{i=0}^{m} [G_v(\theta_i) - \hat{\omega}_{0,i}]^2
\label{eq:velocity_loss}
\end{equation}

\begin{equation}
    L_\alpha = \frac{1}{m} \sum_{i=0}^{m} [G_a(\theta_i) - \hat{\alpha}_{0,i}]^2
\label{eq:acceleration_loss}
\end{equation}

\subsection{Physics-based encoder}

The encoder has been specifically designed to incorporate the kinematic equations of motion directly into the deep learning framework. After predicting the angular velocity and angular acceleration, we follow Equation~\ref{eq:kinematic_general} to combine the independent variable $t$ with the kinematic features. Two LSTM layers process each of $\mathbf{\hat{\omega}}\cdot t$ and $\mathbf{\hat{\alpha}}\cdot t^2$. A concatenation layer combines them, followed by a feed-forward network that combines the features, as shown in Figure~\ref{fig:model}. The encoder outputs the prediction for the displacement angle $\delta\theta(t)$.

\subsection{Graph-based decoder}

The graph-based decoder is added for learning the topological movement information between finger joints. We add the current local rotation value $\theta_0$ to the predicted displacement value $\delta\hat{\theta}(t)$ to get the initial prediction for local rotation $\hat{\theta}(t) \in \mathcal{R}^{3J}$. We then reshape this into $\hat{\theta}(t)_0 \in \mathcal{R}^{J \times 3}$ and input to the graph convolution network ($GCN$). The adjacency matrix $A \in \mathcal{R}^{J \times J}$ of the graph follows the finger skeletal structure (refer Figure~\ref{fig:model}) to allow learning of movement within the fingers. Each GCN layer ($l \in [1, L]$) computes,
\begin{equation}
    \hat{\theta}(t)_{l} = GCN_l(\hat{\theta}(t)_{l-1}) = A\hat{\theta}(t)_{l-1}W_l
\end{equation}
where $W_l \in \mathcal{R}^{F \times F'}$ represents the learnable parameters of the $GCN_l$ and $\hat{\theta}(t)_{l} \in \mathcal{R}^{J\times F'}$ is the output of $GCN_l$. We process the $x,y,z$ axes by setting input-features and output-features of the graph-based decoder to $3$. The output of the GCN is $\hat{\theta}(t)_L$, and the final prediction of the finger rotation for time $t$.

\begin{figure*}[!ht]
\centerline{\includegraphics[scale=0.43]{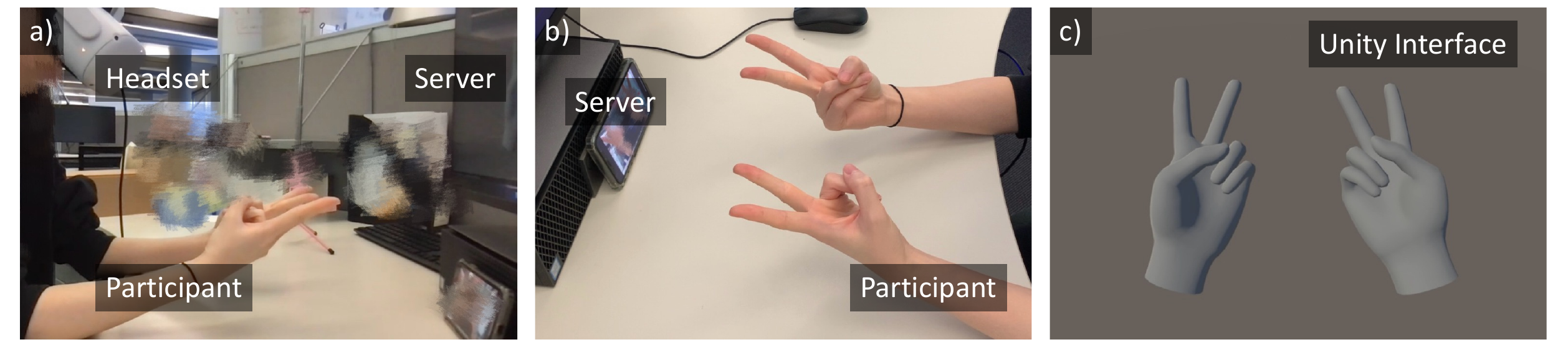}}
\caption{The setup of collecting VRHands dataset through Meta Quest 2 headset: a) the right perspective of the real situation, b) the top perspective of the real situation, and c) the real-time Unity interface when collecting data. }
\label{fig:data-collection}
\end{figure*}

\subsection{Loss function} \label{ssec:loss}

The overall loss metric of this framework is comprised of a cumulative of 12 distinct loss functions. To enforce the time-agnostic weight learning, we use prediction time $t$ as an independent variable in the model. To introduce the model to different prediction times while keeping the training time reasonable, we simultaneously take predictions for $T=[40\text{ms}, 80\text{ms}, ..., 400\text{ms}]$ into the loss metric calculation. Therefore, there are 10 loss functions that follow Equation~\ref{eq:angle_loss} with different prediction times $t \in T$. We introduce 10 weights $w_t \in (0,1]$ for each prediction time $t$ to control the focus of the model optimization during training. Additionally, we have two auxiliary loss functions for angular velocity and angular acceleration as described in Section~\ref{ssec:kfe}. The final loss ($L$) is calculated in Equation~\ref{eq:final_loss}. 

\begin{equation}
    L_\theta(t) = \frac{1}{n} \sum_{i=0}^{n} [\theta(t_0 + t)_i - \hat{\theta}(t_0 + t)_i]^2
\label{eq:angle_loss}
\end{equation}

\begin{equation}
    L = \sum_{t \in T} w_t L_\theta(t) + L_\omega + L_\alpha
\label{eq:final_loss}
\end{equation}


\section{Experiments}\label{sec:exp}

\subsection{Datasets}

We utilized the VRHands dataset that we collected during the study to develop and evaluate the model architecture. The details of this dataset are described in the following paragraph. Additionally, in order to evaluate the generalizability of the proposed architecture, we trained and evaluated the architecture on the Re:InterHand dataset \cite{moon2023reinterhand}, which is one of the latest large-scale hand motion datasets. We utilized the annotations included in the dataset as xyz-coordinates of the corresponding joints. For further details about the Re:InterHand dataset, please refer to the supplementary material.

\paragraph{Data collection:}

We used the Meta Quest 2 headset to provide virtual environment, and Unity software to stream the real-time data. We recruited 7 participants [P1-P7] and utilized a variety of finger movement tasks, including grabbing different objects with different shapes and sizes, abduction and flexion of each finger, writing using a pen, and typing. This ensures generalizability of this dataset to real-world scenarios. Each participant took around 10 minutes to complete all tasks and the data was collected at a sampling rate of 100 Hertz (10 millisecond period). Even though all participants were right-handed, the designed tasks required using both hands simultaneously and data from both hands were captured. The study protocol was approved by the institutional ethics committee. 

\paragraph{Data statistics:}

In the dataset, we collected the interphalangeal data (i.e., the gaps between two bones). Our dataset contains 14 finger joints, and for each joint the rotation was captured around the three axes. We used a naming convention as shown in Figure~\ref{fig:model} left, where $i \in [1,5]$ represents the finger from thumb to little finger, and $j \in [1,3]$ represents the phalanges from palm to fingertip. Since there are three axes $x$, $y$, and $z$, we added suffixes to each joint name. We use finger joint local rotation data as it remains unaffected by the spatial location of the headset and can generalize to different hand sizes. The data is separated into training [P1, P2, P3, P4, P5], validation [P6], and testing [P7] splits. The data collection process is shown in Figure~\ref{fig:data-collection}. Additional statistical analyses of the VRHands dataset, including maximum, mean, and standard deviation values, are detailed in the supplementary materials. 

\subsection{Model parameters}

In the kinematic feature extractor, the moving average filter uses a window length of $10$ and the Gaussian filter has $\sigma=3$. Both LSTMs have $hidden\_dim=32$, $num\_layers=1$, $activation=Tanh(\cdot)$, and both LSTMs are followed by a $(32,1)$ linear layer to reduce the $hidden\_dim$ to $1$. In the physics-based encoder, both LSTMs have $hidden\_dim=32$, $num\_layers=1$, and $activation=Tanh(\cdot)$. The LSTM outputs are concatenated along the $hidden\_dim$ axis to get a new $hidden\_dim=64$. This is fed into a series of $4$ linear layers with weight shapes of $[(64,64), (64,128), (128,32), (32,1)]$. During training, $3$ drop-out layers with $drop\_rate=0.3$ follow first $3$ linear layers. Encoder output for a single step has the shape $(42,1)$. Before giving it to the graph-based decoder, we reshaped it to $(14,3)$ to get $x,y,z$ coordinates of each joint together. In decoder, we used $7$ graph convolution layers with $channels=[(3,32), (32,64), (64,128), (128,128), (128,64), (64,32),$ $(32,1)]$ to extract the information among joints in each finger.

\subsection{Training details}

We implemented TA-GNN with PyTorch 2.0 on a PC with an Intel Core i9-10900K CPU, $64$ GB RAM, and an NVIDIA GeForce RTX 3090 GPU with $24$ GB VRAM. The input to the model is $16$ consecutive data points corresponding to $[-150, -140, ..., -10, 0]$ ms. The model contains $1.49M$ parameters with $9K$ parameters per joint for the kinematic feature extractor, $25K$ parameters per joint for the physics-based encoder, and $37K$ parameters for the decoder. We trained the model $200$ epochs with Adam optimizer \cite{kingma2014adam} with $0.001$ learning rate, $2048$ batch size, $0$ weight decay, and $0.95$ exponential learning rate decay at 20 epoch intervals. We use simultaneous prediction of outputs for $10$ time points to ensure time-agnostic learning as described in Section~\ref{ssec:loss}.


\section{Results} \label{sec:results}

\subsection{Baseline methods}

Previous research has explored hand movement predictions, specifically within the context of virtual environments \cite{chan2007hand,yang2023fully}. However, a notable challenge in comparison with Chan et al. \cite{chan2007hand} was the unavailability of their dataset, their model, and the quantitative evaluation of their results. However, the manuscript presented a graphical representation of errors, detailing the absolute prediction error. We extracted the values for the most optimistic estimates and rounded them to two decimal points using visual inspection. We extracted the values across four different movements in their bar chart and calculated the average errors. To ensure finer comparison, we converted the radians into degrees. Additionally, we compared our results with Yang et al. \cite{yang2023fully} that use fully flexible smart gloves and deep learning motion intention prediction method in VR context. 

We further incorporated the zero-velocity baseline following \cite{martinez2017human} as a simple robust model. The zero-velocity baseline assumes that the most recent observed frame remains constant in subsequent predictions.

\subsection{Evaluation metrics} 

We use two primary evaluation metrics mean-angle-error (MAE) as in Equation~\ref{eq:mae} and mean-squared-error (MSE) as in Equation~\ref{eq:mse} to assess our model and compare with \cite{chan2007hand} and \cite{yang2023fully} respectively. This is because MAE was used in \cite{chan2007hand} and MSE was used in \cite{yang2023fully} separately. We adopted these metrics to enable a direct comparison with the findings of these earlier works. The evaluation metric of the model's performance was conducted quantitatively, using the MAE as a metric to rigorously compare the predicted angles with the empirically established ground truth. We denote the number of samples in the test dataset as $N$ and the number of joints as $J$. Additionally, for Re:InterHand dataset evaluation, we utilize root-mean-square-error (RMSE) in millimetres, between the predicted and ground-truth positions of the joints to evaluate the performance.

\begin{equation}
    \label{eq:mae}
    MAE = \frac{1}{3NJ} \sum_{a=1}^{N} \sum_{i=1}^{3J} \left| \hat{y_{i}}_a - y_{ia} \right|
\end{equation}

\begin{equation}
    \label{eq:mse}
    MSE = \frac{1}{3NJ} \sum_{a=1}^{N} \sum_{i=1}^{3J} (\hat{y_{i}}_a - y_{ia})^2
\end{equation}

\subsection{Quantitative Results}

\begin{table*}[!ht]
    \centering
    \caption{Comparison of MAE results (degrees) for the VRHands dataset.}
    \label{tab:zero-velo}
    \begin{tabularx}{\linewidth}{c|>{\centering\arraybackslash}X>{\centering\arraybackslash}X>{\centering\arraybackslash}X>{\centering\arraybackslash}X>{\centering\arraybackslash}X>{\centering\arraybackslash}X>{\centering\arraybackslash}X>{\centering\arraybackslash}X>{\centering\arraybackslash}X>{\centering\arraybackslash}X}
    \toprule 
    Time (ms) & 40 & 80 & 120 & 160 & 200 & 240 & 280 & 320 & 360 & 400 \\
    \midrule
    \midrule
    TA-GNN & \textbf{0.74} & \textbf{0.98} & \textbf{1.02} & \textbf{1.05} & \textbf{1.09} & \textbf{1.11} & \textbf{1.09} & \textbf{1.16} & \textbf{1.52} & \textbf{2.25}\\
    \midrule
    Zero-velocity & 0.76 & 1.42 & 2.03 & 2.58 & 3.06 & 3.46 & 3.80 & 4.08 & 4.31 & 4.51 \\
    \bottomrule 
    \end{tabularx}
\end{table*}

\begin{table*}[!ht]
    \centering
    \caption{Comparison of time-agnostic MAE results (degrees) for the VRHands dataset.}
    \label{tab:time-agnostic}
    \begin{tabularx}{\linewidth}{c|>{\centering\arraybackslash}X>{\centering\arraybackslash}X>{\centering\arraybackslash}X>{\centering\arraybackslash}X>{\centering\arraybackslash}X>{\centering\arraybackslash}X>{\centering\arraybackslash}X>{\centering\arraybackslash}X>{\centering\arraybackslash}X>{\centering\arraybackslash}X}
    \toprule 
    Time (ms) & 20 & 60 & 100 & 140 & 180 & 220 & 260 & 300 & 340 & 380 \\
    \midrule
    \midrule
    TA-GNN & 0.50 & \textbf{0.90} & \textbf{1.01} & \textbf{1.03} & \textbf{1.07} & \textbf{1.10} & \textbf{1.10} & \textbf{1.10} & \textbf{1.30} & \textbf{1.85}\\
    \midrule
    Zero-velocity & \textbf{0.40} & 1.10 & 1.73 & 2.31 & 2.83 & 3.27 & 3.64 & 3.95 & 4.20 & 4.41\\
    \bottomrule 
    \end{tabularx}
\end{table*}

\begin{table*}[!ht]
    \centering
    \caption{Comparison of RMSE results (mm) for ReInterHands dataset.}
    \label{tab:rmse-newdata}
    \begin{tabularx}{\linewidth}{c|>{\centering\arraybackslash}X>{\centering\arraybackslash}X>{\centering\arraybackslash}X>{\centering\arraybackslash}X>{\centering\arraybackslash}X>{\centering\arraybackslash}X>{\centering\arraybackslash}X>{\centering\arraybackslash}X>{\centering\arraybackslash}X>{\centering\arraybackslash}X>{\centering\arraybackslash}X>{\centering\arraybackslash}X}
    \toprule 
    Time (ms) & 44.44 & 88.88 & 132.32 & 177.76 & 222.20 & 266.64 & 311.08 & 355.52 & 399.96 & 444.4 \\
    \midrule
    \midrule
    TA-GNN & \textbf{6.96} & \textbf{10.15} & \textbf{10.87} & \textbf{10.54} & \textbf{10.23} & \textbf{10.10} & \textbf{10.24} & \textbf{11.16} & \textbf{13.71} & \textbf{17.82} \\
    \midrule
    Zero-velocity & 8.38 & 14.45 & 18.61 & 21.50 & 23.64 & 25.32 & 26.70 & 27.83 & 28.79 & 29.64\\
    \bottomrule
    \end{tabularx}
\end{table*}

\begin{table*}[!ht]
    \centering
    \caption{Comparison of time-agnostic RMSE results (mm) for ReInterHands dataset.}
    \label{tab:time-agnostic-newdata}
    \begin{tabularx}{\linewidth}{c|>{\centering\arraybackslash}X>{\centering\arraybackslash}X>{\centering\arraybackslash}X>{\centering\arraybackslash}X>{\centering\arraybackslash}X>{\centering\arraybackslash}X>{\centering\arraybackslash}X>{\centering\arraybackslash}X>{\centering\arraybackslash}X>{\centering\arraybackslash}X}
    \toprule 
    Time (ms) & 22.22 & 66.66 & 111.10 &  155.54 & 199.98 & 244.42 & 288.86 & 333.30 & 377.74 & 422.18\\
    \midrule
    \midrule
    TA-GNN & \textbf{4.25} & \textbf{8.93} & \textbf{10.73} & \textbf{10.75} & \textbf{10.35} & \textbf{10.13} & \textbf{10.13} & \textbf{9.82} & \textbf{12.26} & \textbf{15.60}\\
    \midrule
    Zero-velocity & 4.50 & 11.70 & 16.73 & 20.18 & 22.64 & 24.53 & 26.04 & 27.00 & 28.32 & 29.23\\
    \bottomrule
    \end{tabularx}
\end{table*}

\begin{table}[!ht]
    \centering
    \caption{Comparisons of MAE in degrees between TA-GNN and Pehm \cite{chan2007hand}. }
    \label{tab:comparison-chan2007}
    \begin{tabularx}{\linewidth}{c|>{\centering\arraybackslash}X>{\centering\arraybackslash}X>{\centering\arraybackslash}X>{\centering\arraybackslash}X}
    \toprule 
    Time (ms) & 100 & 200 & 300 & 400 \\
    \midrule 
    Phm \cite{chan2007hand} & 18.91 & 28.07 & 40.11 & 54.43 \\
    Pehm \cite{chan2007hand} & 13.75 & 20.63 & 28.07 & 36.10 \\
    \midrule 
    Zero-Velocity & 1.84 & 3.27 & 4.24 & 4.81 \\
    \midrule
    TA-GNN & \textbf{1.15} & \textbf{1.37} & \textbf{1.13} & \textbf{2.33} \\
    \bottomrule 
    \end{tabularx}
\end{table}

\begin{table}[!ht]
    \centering
    \caption{Comparisons of MSE between TA-GNN and LSTM-Attention \cite{yang2023fully}. }
    \label{tab:comparison-yang2023}
    \begin{tabularx}{\linewidth}{c|>{\centering\arraybackslash}X>{\centering\arraybackslash}X>{\centering\arraybackslash}X}
    \toprule 
    Time (ms) & 100 & 200 & 400 \\
    \midrule 
    LSTM \cite{yang2023fully} & 203.94 & 422.80 & 523.20 \\
    LSTM-Attention \cite{yang2023fully} & 35.20 & 96.90 & 122.80 \\
    \midrule 
    Zero-Velocity & 45.25 & 109.18 & 183.12 \\
    \midrule
    TA-GNN & \textbf{12.88} & \textbf{17.42} & \textbf{49.96} \\
    \bottomrule 
    \end{tabularx}
\end{table}

In Table \ref{tab:comparison-chan2007}, we compared our results with the hand motion prediction method proposed previously \cite{chan2007hand}. We calculated the average error for all types of movement in their results. All the errors are ranged in $(0, 2)$ radians, which is $(0, 114)$ degrees. TA-GNN shows substantially diminished MAE values in contrast to the other two models. The error commences at 1.15 degrees at 100 ms to 2.33 degrees at 400 ms. The improvement ranges from -88\% to -96\%, indicating a clear reduction in MAE when using TA-GNN. 

In Table \ref{tab:comparison-yang2023}, we compared our results with the fully flexible smart gloves and deep learning motion intention prediction method proposed previously \cite{yang2023fully}. Their method was also proposed in VR context with extra glove sensors as the input. They used 100 data points within 1000 ms before the prediction moment as the input data, while TA-GNN model is able to outperform them while using only 16 data points within 150 ms. TA-GNN shows substantially diminished MSE values in contrast to the two models. The error commences at 12.88 at 100 ms to 49.96 at 400 ms, where the improvement ranges from -59\% to -82\%. 

The MAE for TA-GNN predictions and zero-velocity baseline are presented in Table~\ref{tab:zero-velo} and Table~\ref{tab:time-agnostic}. Compared to the MAE of zero-velocity baseline presented on VRHands dataset shown in Table~\ref{tab:zero-velo}, it is evident that TA-GNN significantly reduces the error, specially for time periods over $120$ ms where the error is reduced to below $50\%$. Table~\ref{tab:time-agnostic} shows the time-agnostic MAE results, where TA-GNN significantly reduces the error when dealing with time periods that it hasn't seen before. 

Furthermore, in order to evaluate the generalizability of the proposed model architecture, we test our model on Re:InterHand dataset \cite{moon2023reinterhand}. The RMSE results in millimetres are shown in Table~\ref{tab:rmse-newdata} and Table~\ref{tab:time-agnostic-newdata}. It is worth noting that our model architecture was able to achieve 45\% improvement on average compared to the zero-velocity baseline. Unfortunately, we were not able to find any previous work that had attempted a similar task with this dataset for comparison.

\subsection{Qualitative Results}

\begin{figure*}[!ht]
\centerline{\includegraphics[scale=0.6]{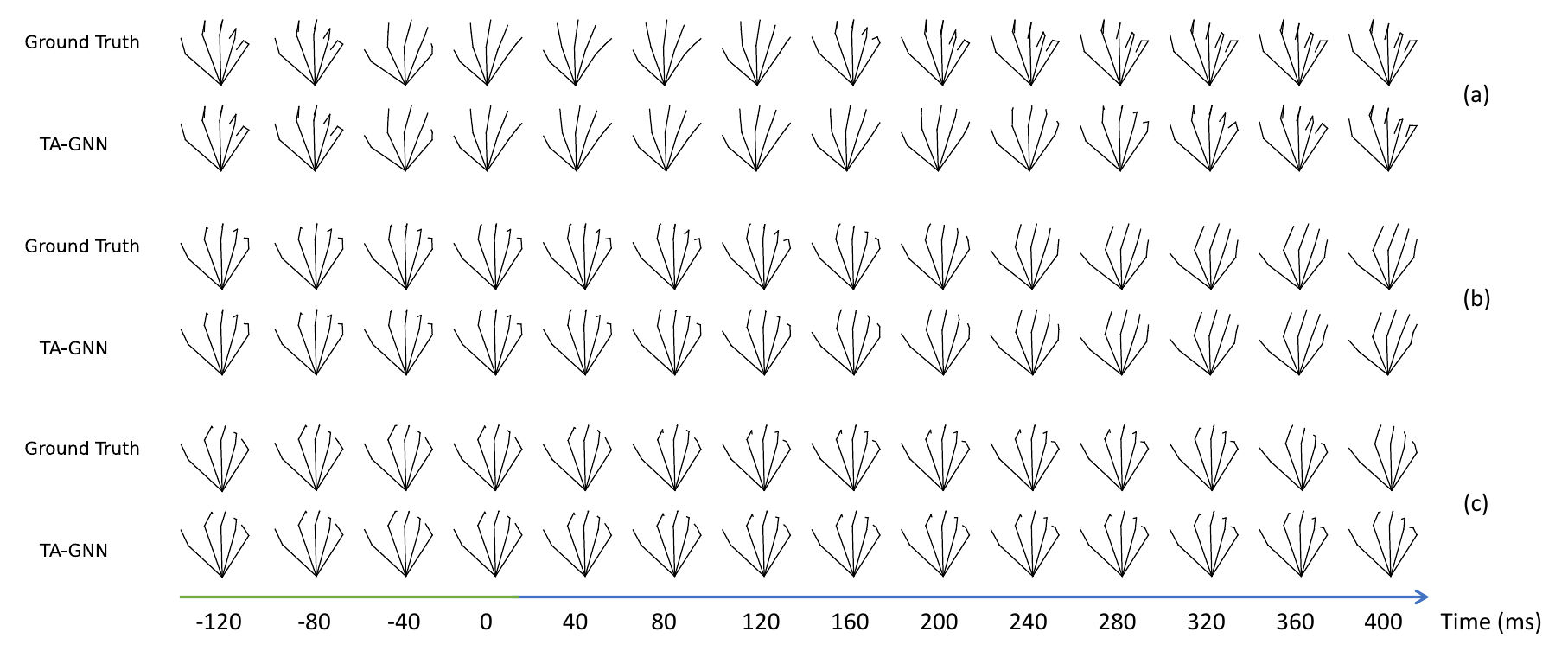}}
\caption{The qualitative results for different models. The time interval is 40 ms. The activities are randomly selected from the movement. The participant (a) made fists, (b) flexed their fingers, and (c) holding an object. }
\label{fig:qualitative-results}
\end{figure*}

Figure~\ref{fig:qualitative-results} presents a qualitative representation of the finger movements predicted by the TA-GNN model compared with the ground truth. In Figure~\ref{fig:qualitative-results} (a), the participant performed the action of forming fists. Figure~\ref{fig:qualitative-results} (b) provides insights into the action of finger flexion. Figure~\ref{fig:qualitative-results} (c) offers an understanding of holding an object. Although the fingers didn't follow the ground-truth in Figure~\ref{fig:qualitative-results} (a) 160 ms - 240 ms, the 400 ms can still predict the exact pose. The TA-GNN predictions exhibit a commendable level of alignment with the ground truth, showcasing the model's adeptness at accurately predicting a variety of motions. We notice that this alignment reduces slightly during the 160 ms - 240 ms time period. However, the model is able to recover and make better predictions for longer times such as 320 ms - 400 ms.

\subsection{Ablation study}

\begin{table*}[!ht]
    \centering
    \caption{Ablation MAE results (degrees) for the TA-GNN model in degrees. KFE: Kinematic-feature-extractor. GCN: Graph-convolutional-network.}
    \label{tab:ablation-table-time-indep}
    \begin{tabularx}{\linewidth}{c|>{\centering\arraybackslash}X>{\centering\arraybackslash}X>{\centering\arraybackslash}X>{\centering\arraybackslash}X>{\centering\arraybackslash}X>{\centering\arraybackslash}X>{\centering\arraybackslash}X>{\centering\arraybackslash}X>{\centering\arraybackslash}X>{\centering\arraybackslash}X}
    \toprule 
    Time (ms) & 40 & 80 & 120 & 160 & 200 & 240 & 280 & 320 & 360 & 400 \\
    \midrule
    \midrule
    TA-GNN & \textbf{0.74} & \textbf{0.98} & \textbf{1.02} & \textbf{1.05} & \textbf{1.09} & \textbf{1.11} & \textbf{1.09} & \textbf{1.16} & \textbf{1.52} & \textbf{2.25} \\
    \midrule
    \midrule
    w/o KFE & 0.93 & 1.20 & 1.37 & 1.48 & 1.51 & 1.43 & 1.30 & 1.32 & 1.65 & 2.34 \\
    & +25.7\% & +22.4\% & +34.3\% & +41.0\% & +38.5\% & +28.8\% & +19.3\% & +13.8\% & +8.6\% & +4.0\% \\
    \midrule
    w/o GCN \& KFE & 0.75 & 1.19 & 1.42 & 1.50 & 1.44 & 1.26 & \textbf{1.09} & 1.22 & 1.72 & 2.47 \\
     & +1.4\% & +21.4\% & +39.2\% & +42.9\% & +32.1\% & +13.5\% & 0.0\% & 5.2\% & 13.2\% & 9.8\%\\
     \midrule
     w/o GCN & 0.79 & 1.27 & 1.51 & 1.58 & 1.50 & 1.32 & 1.17 & 1.32 & 1.82 & 2.53 \\
     & +6.8\% & +29.6\% & +48.0\% & +50.5\% & +37.6\% & +19.0\% & +7.3\% & 13.8\% & 19.7\% & 12.4\% \\
    \midrule
    \midrule
    Zero-velocity & 0.76 & 1.42 & 2.03 & 2.58 & 3.06 & 3.46 & 3.80 & 4.08 & 4.31 & 4.51\\
    & +2.7\% & +44.9\% & +99.0\% & +145.7\% & +180.7\% & +211.7\% & +248.6\% & +251.7\% & +183.6\% & +100.4\% \\
    \bottomrule 
    \end{tabularx}
\end{table*}

We ablate the components of our model to evaluate the contribution of each and select the suitable model parameters. We performed experiments on the time-agnostic property, kinematic feature extractor, physics-based encoder, and the graph-based decoder.

\paragraph{Time agnostic prediction:}

We specifically look at how the model performs on data from time points that it never saw during its training phase. This is done to check if the model can still make accurate predictions that happen at times it hasn't seen before. To evaluate this, we examine the results shown in Table~\ref{tab:time-agnostic} and Table~\ref{tab:time-agnostic-newdata}. These tables demonstrate that, compared to a basic benchmark or standard model, our model is not only capable of handling new, unseen time periods, but it also shows significant improvements in performance. This means that even for time periods that were not part of the training data, the model can predict with a good degree of accuracy, suggesting that it has learned underlying patterns that are not specific to the particular times it was trained on. 

\paragraph{Kinematic feature extractor:} 

We conducted an ablation study to assess the role of the kinematic feature extractor in our model's performance. We removed the learnable components of the kinematic feature extractor, along with the auxiliary losses, and directly fed the approximate kinematic features to the physics-based encoder. The results of this change are shown in Table~\ref{tab:ablation-table-time-indep}, where we observed a notable increase in the MAE, with an average increase of 23\%. The most substantial increase in error occurred at the 160 ms, where the MAE increased 41.0\%. This performance indicates that using a learnable filter along with a method for predicting derivatives on both sides actually leads to better kinematic features being provided to the physics-based model. This enhances the model's ability to predict more accurately, highlighting the importance of these features in achieving high-quality results. 

\paragraph{Physics-based encoder:}


To assess the specific impact of the physics-based encoder within our model, we removed two components: the kinematic feature extractor and the graph-based decoder. This allowed us to isolate and evaluate the performance of the physics-based encoder by itself. According to the data presented in Table~\ref{tab:ablation-table-time-indep}, we observed an average increase in the MAE of 18.0\%. Notably, the most significant improvement occurred at 160 ms, where the increase in MAE reached 42.9\%. These results underline the effectiveness of the physics-based encoder in capturing the temporal kinematic constraints of each joint, despite the model's time-independent design. This significant performance difference, particularly at 160 ms, highlights the essential role of the physics-based encoder in our architecture. 

\paragraph{Graph-based decoder:} 


To thoroughly assess the effectiveness of the graph-based decoder in our model, we conducted an experiment by removing this component. This removal led to an observable degradation in performance, with an overall error increase of 25\% as shown in Table \ref{tab:ablation-table-time-indep}. A deeper examination of the data revealed that the most significant deterioration in accuracy occurred at 160 ms, where the error increased by 50.5\%. This marked the most substantial error increase compared to other parts tested in our ablation study. These results clearly highlight the critical role that the graph-based decoder plays in our model. It significantly enhances the model's ability to process and analyze data efficiently, which is evident from the sharp rise in error following its exclusion.


\section{Discussion}\label{sec:conclusion}

As humans, our brains have evolved to be able to identify natural human motion effortlessly as it is vital for us to interact with each other. Following a similar intuition, TA-GNN tries to utilize the temporal and topological patterns that characterize a natural human motion from a random object motion. It is evident from the results that TA-GNN is able to outperform other methods in predicting future finger movements. Furthermore, TA-GNN is able to make continuous predictions, without the need to train for different prediction times separately. For applications where prediction or rendering has to be done at a high frequency, this ability is critical as training a model with different weights for each prediction time results in a large overhead for the model weights; whereas a time-agnostic model is able to make continuous prediction while being relatively simple and minimal in resources required. This architecture also opens the door for lower frequency measurements during everyday use as the model architecture can be used to synthesize high-frequency data from the low frequency data, which is worth exploring in future work.

Despite the positive results of our research, there are some limitations. The VRHands dataset was collected using Meta Quest 2 headset, which could affect the generalizability of data from other headsets. However, we collected the data from various participants with different backgrounds, ensuring the diversity of our dataset. Additionally, the dataset contains finger angles with respect to the wrist. Therefore, the model is not able to predict the location of the whole hand in space. In the future, we expect to use other sensors and headsets such as HTC Vive and Meta Quest 3 to enhance the data accuracy and make the dataset more comprehensive. Additionally, we aim to expand TA-GNN predictions for arm, upper-body, or whole-body motions to improve its versatility. Furthermore, as fingers movements are quicker relative to larger body parts, we focused on short-term predictions up to 400 ms. In the future, it is worth exploring the model's capability in long-term finger motion prediction.

\section{Conclusion}
We propose a novel physics-based time-agnostic deep-learning model to predict finger movements from historical joint angles, which we call physics-inspired time-agnostic graph neural network (TA-GNN). The proposed model comprises a kinematic feature extractor used to generate angular velocity and angular acceleration from historical movement data. This is followed by a physics-based encoder based on kinematic equations to predict the future displacement angle of each joint separately. The graph-based decoder is used to learn the topological movement information between finger joints. Designed with a time-agnostic approach, the model decouples its weights from designated prediction time periods. This allows a single model to predict future motions up to 400 ms ahead without training for each time period separately. Compared with the baseline methods, TA-GNN was able to significantly reduce the finger motion prediction error. This novel approach enhances finger tracking without additional sensors, enabling predictive interactions such as haptic re-targeting and improving predictive rendering quality.

\begin{acks}
Dr. Withana is a recipient of an Australian Research Council Discovery Early Career Award (DECRA) - DE200100479 funded by the Australian Government. We thank our participants for their valuable time. We appreciate the members of the AID-LAB for assisting us in various ways. 
\end{acks}

\bibliographystyle{ACM-Reference-Format}
\bibliography{bibliography}

\appendix
\section{Re:InterHand Dataset}

Re:InterHand dataset \cite{moon2023reinterhand} contains videos and annotations of interacting hands and therefore poses a realistic challenge in predicting the motion of fingers. The dataset is captured at 90Hz and the videos are semi-automatically annotated with joint positions. We utilized the 'Keypoints' for both left and right hands from the orig\_fits split of the annotation section of the dataset as 3D xyz-coordinates of the joints. We use 21 out of the 22 joints for each hand excluding 'Forearm\_Stub' joint as it is outside the wrist area. We utilized the same model training and testing procedure for this dataset except making adjustments for the difference in number of joints and the difference in frequency of the data capture.

\section{Study Dataset Description} \label{sec:dataset}

In the dataset, we collected the interphalangeal data (i.e., the gaps between two bones). Our dataset contains 14 finger joints, and for each joint the rotation was captured around the three axes. We used a naming convention as shown in \ref{fig:model} left, where $i \in [1,5]$ represents the finger from thumb to little finger, and $j \in [1,3]$ represents the phalanges from palm to fingertip. Since there are three axes $x$, $y$, and $z$, we added suffixes to the end of each joint name for differentiation. We use finger joint local rotation data as it remains unaffected by the spatial location of the headset and can generalize to different hand sizes. We collected data in various heights (160cm - 180cm), weights (40kg - 80kg), and genders (male, female), ensuring the movements captured provided a holistic representation of possible joint movements. Furthermore, our tasks were designed to simultaneously target both hands. The dataset contains over 2 hours of finger motion, with over 750K data points. Most of the joint movements are observed in z-axis (detailed statistics are listed in appendix) as human fingers have most freedom around that axis. X-axis has low movements as it corresponds to twisting of the fingers. In the y-axis metacarpophalangeal ($J_{i1}y$), it shows larger movements corresponding with the ability to spread the fingers. The proximal interphalangeal ($Ji2_z$) moves larger than the distal interphalangeal ($Ji3_z$) and metacarpophalangeal ($Ji1_z$). We report maximum, mean, and standard deviation of the joint rotation for each time period in \ref{tab:dataset-time} and joint rotation around $z$ axis for each phalanges at $400$ ms in \ref{tab:dataset-joint}. The thumb is reported separately as it has different dynamics compared to other four fingers. We can see that the range of the thumb is smaller than the other four fingers, and the distal interphalangeal ($Ji3_z$) moves smaller than the proximal interphalangeal ($Ji2_z$) and metacarpophalangeal ($Ji1_z$). Since the minimum rotation value is always 0, which means the finger is not moving, we did not include it in both tables. Further statistics of the dataset are included in the supplementary. We will release this dataset to facilitate future research in finger motion prediction.

In Unity, we collected the data using local Euler angles in degrees, which means it is relative to the parent transform's rotation. Euler angles enable the representation of a three-dimensional rotation through three distinct rotations around individual axes. In Unity, these rotations occur sequentially around the Z, X, and Y axis. 


Here we add more statistical analysis of the finger motion capture dataset. \ref{tab:dataset} shows the statistical results for every joint and its corresponding axis separately. There are 15 different joints measured, labeled from $J11$ to $J53$ with corresponding x, y, and z axis. For each joint, three statistical metrics are given: maximum (Max), mean, and standard deviation (STD). $J51_z$ corresponding to the metacarpophalangeal of little-finger shows the highest maximum value, which is $120.84$ degrees. The STD reflects the variability of the movement. A high standard deviation, like $25.62$ for joint $J32_z$, suggests a wide range of motion during the capture period, while a low standard deviation indicates consistent, repeated motion.

As expected, the results follow the intuitive movement of our hands. Most of the joint movements are showed in z-axis (\ref{tab:dataset}) as human fingers have most freedom around that axis. x-axis has low movements as it corresponds to twisting of the fingers. In the y-axis metacarpophalangeal ($J_{i1}y$) shows larger movements corresponding with the ability to spread the fingers. The proximal interphalangeal ($Ji2_z$) moves larger than the distal interphalangeal ($Ji3_z$) and metacarpophalangeal ($Ji1_z$).

\begin{table}[!ht]
    \centering
    \caption{Statistics of the VRHands dataset on different finger joint locations on 400ms. }
    \label{tab:dataset-joint}
    \begin{tabularx}{\linewidth}{c|>{\centering\arraybackslash}X>{\centering\arraybackslash}X|>{\centering\arraybackslash}X>{\centering\arraybackslash}X>{\centering\arraybackslash}X}
    \toprule 
    Joint Location & $J11_z$ & $J12_z$ & $Ji1_z$ & $Ji2_z$ & $Ji3_z$ \\
    \midrule 
    Max & 58.58 & 58.15 & 100.22 & 103.84 & 68.62 \\ 
    Mean (MAE) & 4.21 & 5.73 & 16.65 & 16.99 & 10.40 \\
    STD (RMSE) & 5.71 & 7.76 & 20.39 & 23.59 & 14.50 \\
    \bottomrule 
    \end{tabularx}
\end{table}

\begin{table*}[!ht]
    \centering
    \caption{Statistics of the VRHands dataset for all finger joints on 400ms. }
    \label{tab:dataset}
    \begin{tabularx}{\linewidth}{c|>{\centering\arraybackslash}X>{\centering\arraybackslash}X>{\centering\arraybackslash}X>{\centering\arraybackslash}X>{\centering\arraybackslash}X>{\centering\arraybackslash}X>{\centering\arraybackslash}X>{\centering\arraybackslash}X>{\centering\arraybackslash}X>{\centering\arraybackslash}X>{\centering\arraybackslash}X>{\centering\arraybackslash}X>{\centering\arraybackslash}X>{\centering\arraybackslash}X}
    \toprule 
    Joint & $J11_x$ & $J12_x$ & $J21_x$ & $J22_x$ & $J23_x$ & $J31_x$ & $J32_x$ & $J33_x$ & $J41_x$ & $J42_x$ & $J43_x$ & $J51_x$ & $J52_x$ & $J53_x$ \\
    \midrule 
    Max & 3.19 & 4.39 & 2.77 & 0.11 & 1.21 & 0.73 & 0.12 & 5.85 & 1.04 & 1.26 & 3.75 & 3.93 & 2.64 & 2.33 \\
    Mean & 0.27 & 0.45 & 0.44 & 0.02 & 0.18 & 0.11 & 0.02 & 0.99 & 0.15 & 0.22 & 0.64 & 0.71 & 0.45 & 0.38 \\
    STD & 0.34 & 0.60 & 0.52 & 0.02 & 0.25 & 0.12 & 0.03 & 1.33 & 0.17 & 0.31 & 0.85 & 0.87 & 0.62 & 0.54 \\
    \bottomrule 
    \toprule 
    Joint & $J11_y$ & $J12_y$ & $J21_y$ & $J22_y$ & $J23_y$ & $J31_y$ & $J32_y$ & $J33_y$ & $J41_y$ & $J42_y$ & $J43_y$ & $J51_y$ & $J52_y$ & $J53_y$ \\
    \midrule 
    Max & 2.21 & 2.11 & 25.67 & 0.15 & 0.85 & 10.39 & 0.15 & 3.43 & 9.10 & 1.67 & 2.91 & 29.23 & 3.25 & 1.68 \\
    Mean & 0.09 & 0.13 & 3.30 & 0.02 & 0.09 & 1.48 & 0.03 & 0.47 & 1.29 & 0.28 & 0.38 & 3.12 & 0.47 & 0.19 \\
    STD & 0.19 & 0.24 & 3.63 & 0.04 & 0.18 & 1.53 & 0.04 & 0.76 & 1.29 & 0.41 & 0.57 & 3.73 & 0.70 & 0.30 \\
    \bottomrule 
    \toprule 
    Joint & $J11_z$ & $J12_z$ & $J21_z$ & $J22_z$ & $J23_z$ & $J31_z$ & $J32_z$ & $J33_z$ & $J41_z$ & $J42_z$ & $J43_z$ & $J51_z$ & $J52_z$ & $J53_z$ \\
    \midrule 
    Max & 58.58 & 58.15 & 83.72 & 104.76 & 74.44 & 97.80 & 105.20 & 73.89 & 98.51 & 101.62 & 63.71 & 120.84 & 103.78 & 62.46 \\
    Mean & 4.21 & 5.73 & 13.69 & 15.98 & 10.00 & 16.05 & 18.72 & 11.81 & 17.60 & 17.68 & 10.08 & 19.27 & 15.59 & 9.70 \\
    STD & 5.71 & 7.76 & 16.94 & 23.19 & 14.63 & 19.17 & 25.62 & 16.06 & 21.43 & 23.99 & 13.58 & 24.04 & 21.55 & 13.71 \\
    \bottomrule 
    \end{tabularx}
\end{table*}

\begin{table*}[!ht]
    \centering
    \caption{Statistics of the VRHands dataset on different time periods. }
    \label{tab:dataset-time}
    \begin{tabularx}{\linewidth}{c|>{\centering\arraybackslash}X>{\centering\arraybackslash}X>{\centering\arraybackslash}X>{\centering\arraybackslash}X>{\centering\arraybackslash}X>{\centering\arraybackslash}X>{\centering\arraybackslash}X>{\centering\arraybackslash}X>{\centering\arraybackslash}X>{\centering\arraybackslash}X>{\centering\arraybackslash}X>{\centering\arraybackslash}X}
    \toprule 
    Time (ms) & 40 & 80 & 100 & 120 & 160 & 200 & 240 & 280 & 300 & 320 & 360 & 400 \\
    \midrule 
    Max & 89.66 & 112.13 & 114.41 & 116.62 & 118.05 & 118.08 & 118.10 & 120.71 & 120.32 & 120.112 & 120.54 & 120.84 \\
    Mean (MAE) & 0.80 & 1.50 & 1.84 & 2.15 & 2.75 & 3.27 & 3.72 & 4.08 & 4.24 & 4.38 & 4.61 & 4.81 \\
    STD (RMSE) & 3.25 & 5.54 & 6.49 & 7.31 & 8.78 & 9.90 & 10.87 & 11.52 & 11.77 & 11.98 & 12.36 & 12.61 \\
    \bottomrule 
    \end{tabularx}
\end{table*}

\section{Model Parameters}

In the kinematic feature extractor, the moving average filter uses a window length of $10$ and the Gaussian filter has $\sigma=3$. Both LSTMs have $hidden\_dim=32$, $num\_layers=1$, $activation=Tanh(\cdot)$, and both LSTMs are followed by a $(32,1)$ linear layer to reduce the $hidden\_dim$ to $1$. In the physics-based encoder, both LSTMs have $hidden\_dim=32$, $num\_layers=1$, and $activation=Tanh(\cdot)$. The LSTM outputs are concatenated along the $hidden\_dim$ axis to get a new $hidden\_dim=64$. This is fed into a series of $4$ linear layers with weight shapes of $[(64,64), (64,128), (128,32), (32,1)]$. During training, $3$ drop-out layers with $drop\_rate=0.3$ follow first $3$ linear layers. Encoder output for a single step has the shape $(42,1)$. Before giving it to the graph-based decoder, we reshaped it to $(14,3)$ to get $x,y,z$ coordinates of each joint together. In decoder, we used $7$ graph convolution layers with $channels=[(3,32), (32,64), (64,128), (128,128), (128,64), (64,32),$ $(32,1)]$ to extract the relationships among joints in each finger.

\section{Qualitative Results} \label{sec:qual}

\begin{figure*}[!ht]
\centerline{\includegraphics[width=\linewidth]{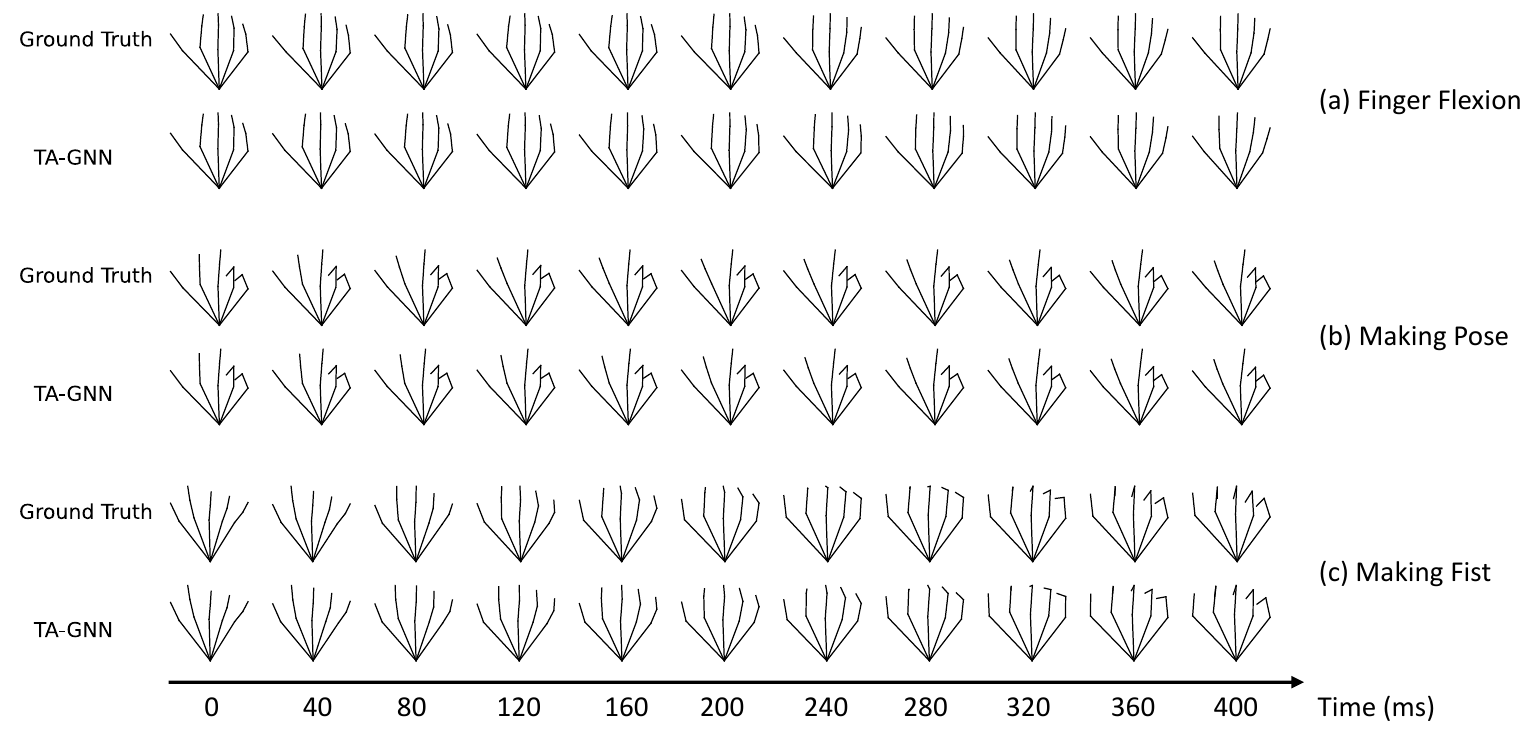}}
\caption{The qualitative results for different models. The activities are randomly selected from the movement. The participant (a) flexed fingers; (b) made pose; (c) made fist; and (d) made pose. }
\label{img:supp-qual}
\end{figure*}

To further demonstrate the effectiveness of TA-GNN, we illustrate more predicted samples in \ref{img:supp-qual}, which contains up to 400 milliseconds predicted results in four different random situations. 

\end{document}